\documentclass[traditabstract,letter,longauth]{aa}
\usepackage{graphicx,natbib,color}
\bibpunct{(}{)}{;}{a}{}{,} 
\usepackage{txfonts,aalongtable}

\newcommand{\Brg}{Br$\gamma\ $}
\newcommand{\Brd}{Br$\delta\ $}

\def\lta{ \lower .75ex\hbox{$\sim$} \llap{\raise .27ex \hbox{$<$}} }

\begin{document}

\date{Received .../Accepted ...}

\title{Accretion-ejection morphology of the microquasar SS433 resolved at sub-au scale\thanks{Based on observations made with VLTI/Gravity instrument.}}
\author{GRAVITY Collaboration\thanks{GRAVITY is developed in a collaboration by the Max Planck Institute for extraterrestrial Physics, LESIA of Paris Observatory / CNRS / UPMC / Univ. Paris Diderot and IPAG of Universit\'e Grenoble Alpes / CNRS, the Max Planck Institute for Astronomy, the University of Cologne, the Centro Multidisciplinar de Astrof\'isica Lisbon and Porto, and the European Southern Observatory.}:
P.-O. Petrucci \inst{1} \and
I.~Waisberg \inst{2} \and
J.-B.~Le~Bouquin \inst{1} \and
J.~Dexter \inst{2} \and
G.~Dubus \inst{1} \and
K.~Perraut \inst{1} \and
P.~Kervella \inst{3,4} \and
R.~Abuter \inst{5} \and
A.~Amorim \inst{6} \and
N.~Anugu \inst{7} \and
J.P.~Berger \inst{5,1} \and
N.~Blind \inst{8} \and
H.~Bonnet \inst{5} \and
W.~Brandner \inst{9} \and
A.~Buron \inst{2} \and
\'E.~Choquet \inst{3,10}\and
Y.~Cl\'enet \inst{3} \and
W.~de~Wit \inst{11} \and
C.~Deen \inst{2} \and
A.~Eckart \inst{12,13} \and
F.~Eisenhauer \inst{2} \and
G.~Finger \inst{5} \and
P.~Garcia \inst{7} \and
R.~Garcia Lopez \inst{9} \and
E.~Gendron \inst{3} \and
R.~Genzel \inst{2,14} \and
S.~Gillessen \inst{2} \and
F.~Gonte \inst{5} \and
X.~Haubois \inst{11} \and
M.~Haug \inst{2,5} \and
F.~Haussmann \inst{2} \and
Th.~Henning \inst{9} \and
S.~Hippler \inst{9} \and
M.~Horrobin \inst{12} \and
Z.~Hubert \inst{9,3} \and
L.~Jochum \inst{5} \and
L.~Jocou \inst{1} \and
Y.~Kok \inst{2} \and
J.~Kolb \inst{11} \and
M.~Kulas \inst{9} \and
S.~Lacour \inst{3} \and
B.~Lazareff \inst{1} \and
P.~L\`ena \inst{3} \and
M.~Lippa \inst{2} \and
A.~M\'erand \inst{5} \and
E.~M\"uller \inst{9,11} \and
T.~Ott \inst{2} \and
J.~Panduro \inst{9} \and
T.~Paumard \inst{3} \and
G.~Perrin \inst{3} \and
O.~Pfuhl \inst{2} \and
J.~Ramos \inst{9} \and
C.~Rau\inst{2} \and
R.-R.~Rohloff \inst{9} \and
G.~Rousset \inst{3} \and
J.~Sanchez-Bermudez \inst{9} \and
S.~Scheithauer \inst{9} \and
M.~Sch\"oller \inst{5} \and
C.~Straubmeier \inst{12} \and
E.~Sturm \inst{2} \and
F.~Vincent \inst{3}\and
I.~Wank \inst{12} \and
E.~Wieprecht \inst{2} \and
M.~Wiest \inst{12} \and
E.~Wiezorrek \inst{2} \and
M.~Wittkowski \inst{5} \and
J.~Woillez \inst{5} \and
S.~Yazici \inst{2,12} \and
G.~Zins \inst{11}
}

\institute{ 
                 Univ. Grenoble Alpes, CNRS, IPAG, F-38000 Grenoble, France
         \and
                 Max Planck Institute for extraterrestrial Physics, Giessenbachstr., 85748 Garching, Germany
         \and
                 LESIA, Observatoire de Paris, PSL Research University, CNRS, Sorbonne Universit\'es, UPMC Univ. Paris 06, Univ. Paris Diderot, Sorbonne Paris Cit\'e, France
         \and
  Unidad Mixta Internacional Franco-Chilena de Astronom\'ia (CNRS UMI 3386), Departamento de Astronom\'ia, Universidad de Chile, Camino El Observatorio 1515, Las Condes, Santiago, Chile
         \and
European Southern Observatory, Karl-Schwarzschild-Str. 2, 85748 Garching, Germany
         \and
CENTRA and Universidade de Lisboa - Faculdade de Ci\^encias, Campo Grande, 1749-016 Lisboa, Portugal
         \and
                 CENTRA and Universidade do Porto - Faculdade de Engenharia,  4200-465 Porto, Portugal
         \and
Observatoire de Gen\`eve, Universit\'e de Gen\`eve, 51 ch. des Maillettes, 1290 Versoix, Switzerland
         \and  
Max-Planck-Institut f\"ur Astronomie, K\"onigstuhl 17, 69117 Heidelberg, Germany
	\and
Hubble Fellow, Jet Propulsion Laboratory, California Institute of Technology, 4800 Oak Grove Drive, Pasadena, CA 91109, USA
         \and
 European Southern Observatory, Casilla 19001, Santiago 19, Chile
         \and
1. Physikalisches Institut, Universit\"at zu K\"oln, Z\"ulpicher Str. 77, 50937 K\"oln, Germany
         \and
Max-Planck-Institute for Radio Astronomy, Auf dem H\"ugel 69, 53121 Bonn, Germany
         \and     
Department of Physics, Le Conte Hall, University of California, Berkeley, CA 94720, USA
   }
 
%

\abstract{We present the first optical observation at sub-milliarcsecond (mas) scale of the microquasar SS 433 obtained with the GRAVITY instrument on the VLT interferometer. The 3.5 hour exposure reveals a rich K-band spectrum dominated by hydrogen \Brg and  \ion{He}{i} lines, as well as (red-shifted) emission lines coming from the jets. The K-band continuum emitting region is dominated by a marginally resolved point source ($<$ 1 mas) embedded inside a diffuse background accounting for 10\% of the total flux. The jet line positions agree well with the ones expected from the jet kinematic model, an interpretation also supported by the  consistent sign (i.e. negative/positive for the receding/approaching jet component) of the phase shifts observed in the lines.  The significant visibility drop across the jet lines, together with the small and nearly identical phases for all baselines, point toward a jet that is offset by less than 0.5 mas from the continuum source and resolved in the direction of propagation, with a typical size of 2 mas. The jet position angle of $\sim$80$^{\circ}$ is consistent with the expected one at the observation date. Jet emission so close to the central binary system would suggest that line locking, if relevant to explain the amplitude and stability of the 0.26c jet velocity, operates on elements heavier than hydrogen. The \Brg profile is broad and double peaked. It is better resolved than the continuum and the change of the phase signal sign across the line on all baselines suggests an East-West oriented geometry alike the jet direction and supporting a (polar) disk wind origin. 
}


\maketitle

\section{Introduction}
SS 433 (K=8.2) is a very well known microquasar discovered in the 1970s 
\citep{ste77}. It is an eclipsing X-ray binary system, the primary component being likely a black hole of $\sim$ 5-16 $M_{\sun}$ (e.g. \citealt{che13} and references therein) accreting matter from a companion massive star (e.g. \citealt{gie02,hil08}). It is located at 5.5 $\pm$ 0.2 kpc from the Sun (e.g. \citealt{blu04}) and has an orbital period of 13.1 days \citep{fabr04}. It is one of the most persistent sources of relativistic jets in our galaxy. Embedded in the radio nebula W 50, it possesses a double-sided radio jet precessing on a period of 162.5 days and tracing out a cone of polar angle $\sim$20$^{\circ}$ around a precession axis of position angle PA$\sim$98$^{\circ}$. 


The jet of SS 433 has been intensively studied from arcminutes down to milliarcsecond (mas) scale in the radio. VLBI observations show that jet signatures are already present at about 2 mas from the center \citep{par99}. SS 433 is the only microquasar  whose jet is also observable in optical emission lines. The extremely large Doppler shifts seen in the optical lines are interpreted as the direct signature of jet material expelled at $\sim 0.26 c$ \citep{eik01}.   
The powerful and continuous accretion flow is provided by the donor star at a rate of $\sim 10^{-4}$ $M_{\sun}/yr$, forming a complex structure around the system. Orbital phase-resolved optical spectra show evidence of an accretion disc in the inner region, which is also thought to power the ejection of matter (e.g. \citealt{per09}). The accretion disc is not directly observable due to the presence of a dense wind outflowing from the disc itself 
 but was revealed by the detection of a pair of widely-separated, hence rapidly rotating ($\sim$ 500 km s$^{-1}$), narrow components in the profile of the "stationary" (i.e. not associated with the jet) Br$\gamma$ and H$\alpha$ lines \citep{per09,per10}. The disc is expected to be perpendicular to the jet axis and precessing with it at the same period (e.g. \citealt{blu11}). The IR spectrum of SS 433 is compatible with bremsstrahlung emission produced by the accretion disc wind, the estimated size at 2 microns of the accretion disc + wind system then being smaller than 1 mas \citep{fuc06}.
 

We present in this letter the observation of SS 433 in the K band performed in July 2016 with the VLTI/GRAVITY instrument. The spectro-interferometric capabilities of GRAVITY  allow us to resolve, for the first time in the Optical, this microquasar at sub-mas spatial resolution. This gives us the opportunity to  study on such scales and simultaneously the properties of the different accretion-ejection components (jets, wind, disk), providing a new look at this famous source. 


\section{Observations and data reduction}
\begin{figure}[!t]
\begin{center}
\includegraphics[width=0.9\columnwidth]{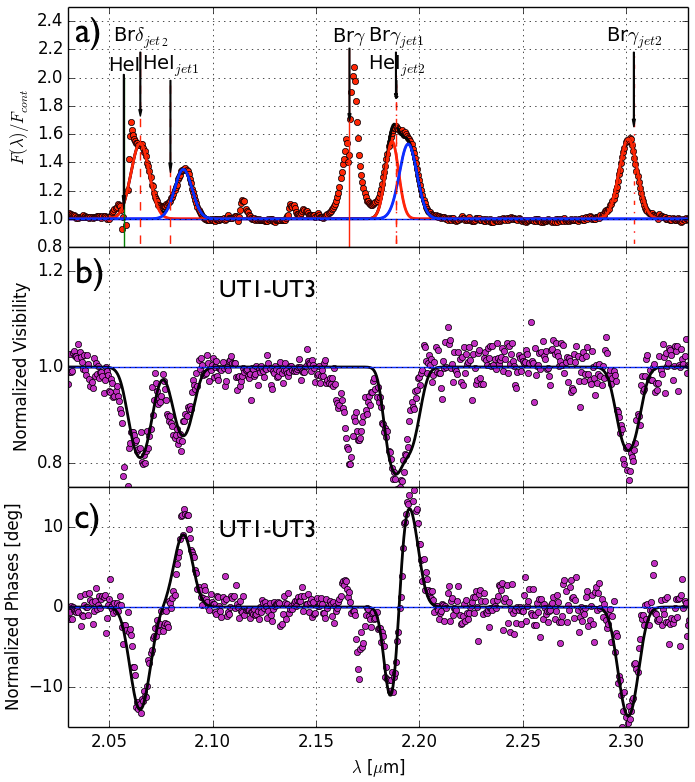}
 \caption{a) Normalized K-band spectrum (with a binning over 2 spectral channels) obtained with GRAVITY. The \Brg and \ion{He}{i} lines as well as the expected position of the corresponding jet lines ($jet1/jet2$ for the approaching/receding jet) are indicated by vertical lines. We have also reported the expected position of the Br$\delta_{jet2}$ line from the receding jet. The solid blue/red line corresponds to the emission of the approaching/receding jet components. b) Visibility amplitudes and c) phases on the UT1-UT3 baseline (the visibility amplitudes and phases of all the baselines as well as the $uv$-plane at the time of the observation are shown in the Appendix, in Fig. \ref{figtotapp}). For the phase, we follow the sign convention of \cite{pau05}, i.e. negative phases point to the baseline direction. The solid black line corresponds to the best fit model of the jet lines (see Sect. \ref{jetline}).
 }
\label{fig1}
\end{center}
\end{figure}
We used GRAVITY (GRAVITY collaboration, 2017, submitted; \citealt{eis11}) to perform spectral-differential interferometry on SS 433 with the Unit Telescopes (UTs) of the VLTI. We recorded  18 data sets at high spectral resolution ($R=4000$) over about 3.5 hours on 2016, July 17th. Each measurement consists of 5 exposures of 60 s integration time. We interlaced our observations with recordings on two interferometric calibrators HD\ 175322 and HD\ 181414\footnote{The "uniform disk" diameters in the K band are 0.214 $\pm$ 0.015 mas and 0.143 $\pm$ 0.010 mas for HD\ 175322 (Sp. type F7V) and HD\ 181414 (Sp. type A2V) respectively.}. We used the GRAVITY standard pipeline \citep{lap14} to reduce and calibrate the observations. We have checked that the different files are similar within errors. Then, to increase the signal-to-noise ratio, we merged all the files and binned over 2 spectral channels the interferometric observables. We used the interferometric data provided by the fringe tracker (working at a higher frequency, i.e. around 900\ Hz) to calibrate the continuum. The $uv$-plane at the date of observation is indicated in the Appendix (Fig. \ref{figtotapp}).


\section{Results}

\subsection{K-band spectrum}
\label{sectSpec}
The K-band GRAVITY spectrum is plotted on the top panel of Fig. \ref{fig1}. It is very rich with hydrogen \Brg and  \ion{He}{i} lines, as well as broad features around 2.08, 2.18 and 2.3 $\mu$m. The stationary  \Brg line is clearly double peaked (see also Fig. \ref{Brg}) while the stationary He I line shows a P\,Cygni profile arising in the wind. We have overplotted in Fig. \ref{fig1} the positions of the  \ion{He}{i}, Br$\gamma$ and  Br$\delta$ approaching (hereafter jet1, pointing to East) and receding (hereafter jet2, pointing to West) jet lines as predicted by the standard kinematic model formula (see \citealt{fab04}, hereafter F04, and reference therein):
\begin{equation}
1+z^{\pm}=\gamma (1\pm\beta\sin\theta\sin i\cos\phi\pm\beta\cos\theta\cos i)
\end{equation}
where the - and + signs correspond to the jet1 and jet2 component respectively, $\beta$ the jet velocity in unit of light speed, $\theta$ the precession angle between the jets and the precessional axis, $i$ the angle between the precessional axis and the line of sight and $\psi$ the precessional phase at the observation date. We use the values determined by \cite{eik01} for $\beta$, $\theta$ and $i$ i.e. $\beta$ = 0.2647, $\theta$ = 20.92$^{\circ}$ and $i$ = 78.05$^{\circ}$. At the date of the GRAVITY observation, the estimated precessional phase is $\psi\sim 0.71$. The jet line positions computed with these parameter values  (vertical dashed line in Fig. 1) 
agree quite well with the position of the observed emission features, clearly supporting a jet origin. The fact that the signs of the phase shifts (Fig. \ref{fig1}c, see below) are the same for all the "jet1" lines and all the "jet2" lines, i.e.  positive (negative) for  jet1 (jet2), also supports that the spatially resolved lines originate in the jets. These features will be called "jet lines" in the following. The parameters of our jet lines fit (redshift, FWHM, equivalent width) are reported in Appendix (Table \ref{tab1}). 

Note that at the precession phase of $\sim$0.71, the jet is almost in the plane of the sky (a favorable geometry to resolve the jet extension) and, due to the transverse Doppler effect, the receding and approaching shifts z$^{\pm}$ are both redshifts. Moreover, the approaching  \ion{He}{i} and receding Br$\gamma$ lines are very close and not distinguishable, producing a blended emission line profile around 2.18 $\mu$m. This precessional phase also corresponds to a near edge-on disk orientation, a precessional phase where P\,Cygni profiles are commonly observed in H and He I lines (F04).\\

\subsection{Visibilities and Phases}
\label{sectInterf}
\subsubsection{Continuum}
The absolute visibility amplitudes of the continuum are all higher than $0.7$ and display a systematic drop versus baseline length. No closure phase is measurable in the continuum. A simple modeling with a Gaussian disk of the K-band continuum emitting region shows that it is dominated by a marginally resolved source (typical size $\sim$0.8 mas) embedded inside a diffuse background accounting for 10\% of the total flux. 

In the following, we normalize the visibility amplitudes and phases by their values in the continuum in order to work with differential quantities. Consequently the \emph{measured} visibility amplitudes in the continuum (taken from the fringe tracker), hereafter called $V_c$, are included into the modeling. As an example, the normalized visibility amplitudes and phases for the UT1-UT3 baseline are shown in the middle and at the bottom of Fig. \ref{fig1} respectively (the normalized visibility amplitudes and phases of the six baselines are plotted in the Appendix in Fig. \ref{figtotapp}).

\subsubsection{Jet lines}
\label{jetline}
We decompose the jet lines contribution into two parts, $F_{\rm jet1}(\lambda)$ /$F_{\rm jet2}(\lambda)$ from the jet1 (approaching)/jet2 (receding) components separately as detailed in Appendix \ref{appA}. 
Looking at the visibility and phases, we infer that jet1 and jet2 are roughly symmetric around the continuum and that the \Brg and He I lines are emitted from the same regions. Consequently the interferometric differential observables can be modeled with:
\begin{equation}
V_{\rm Norm}(u,v,\lambda) =\frac{F_{\rm jet1}(\lambda) V(u,v) + F_{\rm jet2}(\lambda) V(-u,-v) + V_c(u,v)}{V_c(u,v)[F_{\rm jet2}(\lambda) + F_{\rm jet1}(\lambda) + 1]}
\label{eqmodel}
\end{equation}
where $V(u,v)$ is the Fourier Transform of the geometrical model for Jet1.

The significant visibility drop across the lines (with deeper drops on the longest baselines), together with the significant ($>$ 10$^\circ$ for each spectral line) and nearly identical phases for all baselines, point toward an overall jet geometry which is significantly resolved, with a typical size of 2 mas and only slightly offset from the continuum, typically by less than 1 mas.

This is confirmed by tentative adjustments with components located further away (one blob, several blobs, varying PA, flux ratio, elongation, intensity profile...). Secondary minima exist in the $\chi^2_r$ cube for solutions located further away (typically 10\,mas with PA$\approx$70\,deg and 15\,mas with PA$\approx$90\,deg). However the fit qualities are low ($\chi^2_r>3$) and they predict a wrong sign of the visibility phase for at least one baseline. In fact, fitting all jet lines together under the assumption that they have similar origins and structures, makes the fit rather well constrained because of the additional spatial frequency coverage. We conclude that elongated models ($>$ 2.5 mas) for the jet are disfavored with the present dataset.\\

A model composed of a single resolved and slightly offset Gaussian provides a statistically better fit ($\chi^2_r=1.36$). An elongated (along the jet axis) Gaussian blob gives a statistically similar fit ($\chi^2$=1.34). We decided to explore a more abrupt intensity profile for the jet:
\begin{equation}
V(u,v,s,a) = \frac{\exp(2i\pi f a)}{1+2i\pi f s}
\end{equation}
with $f = v \cos(PA) - u \sin(PA)$. This is the Fourier Transform of an exponentially decreasing profile $\exp(-(r-a)/s)\,H(r-a)$, where $H(r)$ is the Heaviside function. It is limited to the positive ordinates, decays on a spatial scale $s$ and is translated by $a$ from the continuum position. Positive $r$ are defined toward PA (North to East), and the model is infinitely thin in the direction perpendicular to PA. The fit is significantly better ($\chi^2_r$=0.89). The best fit parameter values (with their 3$\sigma$ errors\footnote{To compute the errors, we assume 60 observables (6 baselines $\times$ 3 lines (\Brg, \Brd and He I) $\times$ 2 jet components (only 1 for \Brd) $\times$ 2 interferometric measurements (visibility amplitude and phase))  and 3 model parameters ($a$,$s$ and $PA$), giving 57 degrees of freedom}) are $s=1.7\pm0.6$ mas, $a=-0.15\pm0.34$ mas and PA=75$\pm$20$^{\circ}$. The resulting best fit is reported in Fig. \ref{fig1} in black solid line. We tested the thickness of the jet in the transverse direction by convolving the intensity profile with a gaussian. The jet profile is unresolved in the transverse direction, with an upper limit of its transverse size of $\sim$1.2 mas at 3$\sigma$. This upper limit is quite large however due to the coincidental alignment of the baselines with the jet PA (see the $uv$-plane in the Appendix in Fig. \ref{figtotapp}).

\subsubsection{Stationary lines: \Brg}
\begin{figure}[!t]
\includegraphics[width=0.95\columnwidth]{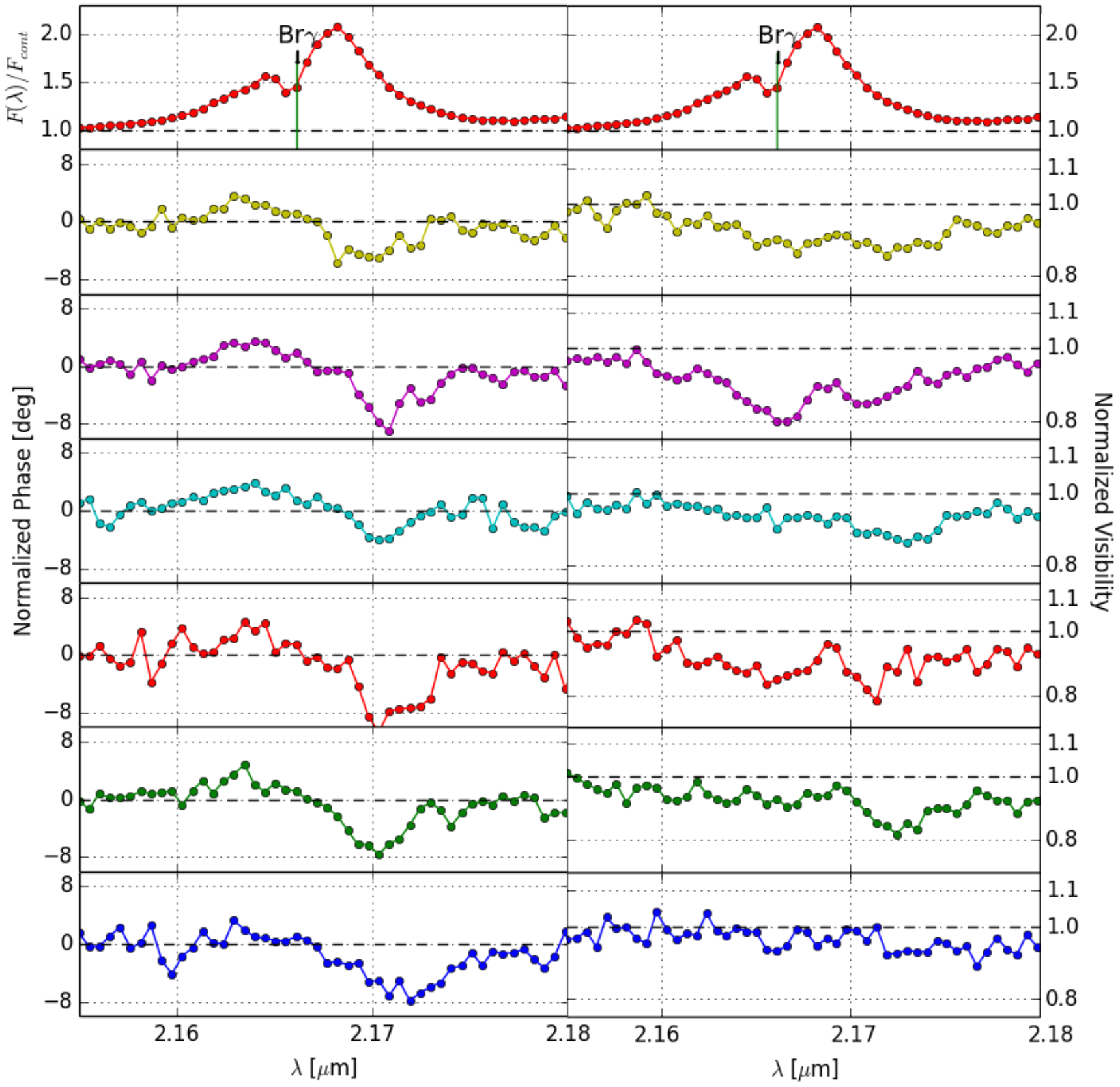}  
\caption{Normalized spectrum (top), visibility phases (left) and amplitudes (right) around the \Brg line with a spectral binning of 2. The symbol colors correspond to the baseline colors indicated in Fig. \ref{figtotapp}.}
\label{Brg}
\end{figure}

The \Brg profile is broad and double peaked. 
The visibilities clearly drop across the line for all the baselines, with a decrease that can reach 20\% with respect to the continuum (Fig. \ref{Brg}). The drop is also deeper for longer baselines. The emitting region size is found to be $\sim$1 mas, similar to the extension of the jets and a little larger than the continuum. The phases in the red part are significantly larger (in absolute values) than in the blue part.
They show a similar behavior for all the baselines, starting from positive values of a few degrees in the blue part of the line, and going to negative values in the red part (Fig. \ref{Brg}). This suggests an East-West oriented geometry, i.e., in a direction similar to the jet one.  
This East-West direction disfavors an origin from the accretion disk, which is expected to be perpendicular to the jet axis (i.e. South-North). A (rather polar) disk wind is a more natural candidate. The change of the phase sign across the line can be explained then by the presence of an approaching (positive phase) and a receding (negative phase) component in the wind. 

\section{Discussion}
\begin{figure}[!t]
\begin{center}
\includegraphics[width=0.9\columnwidth]{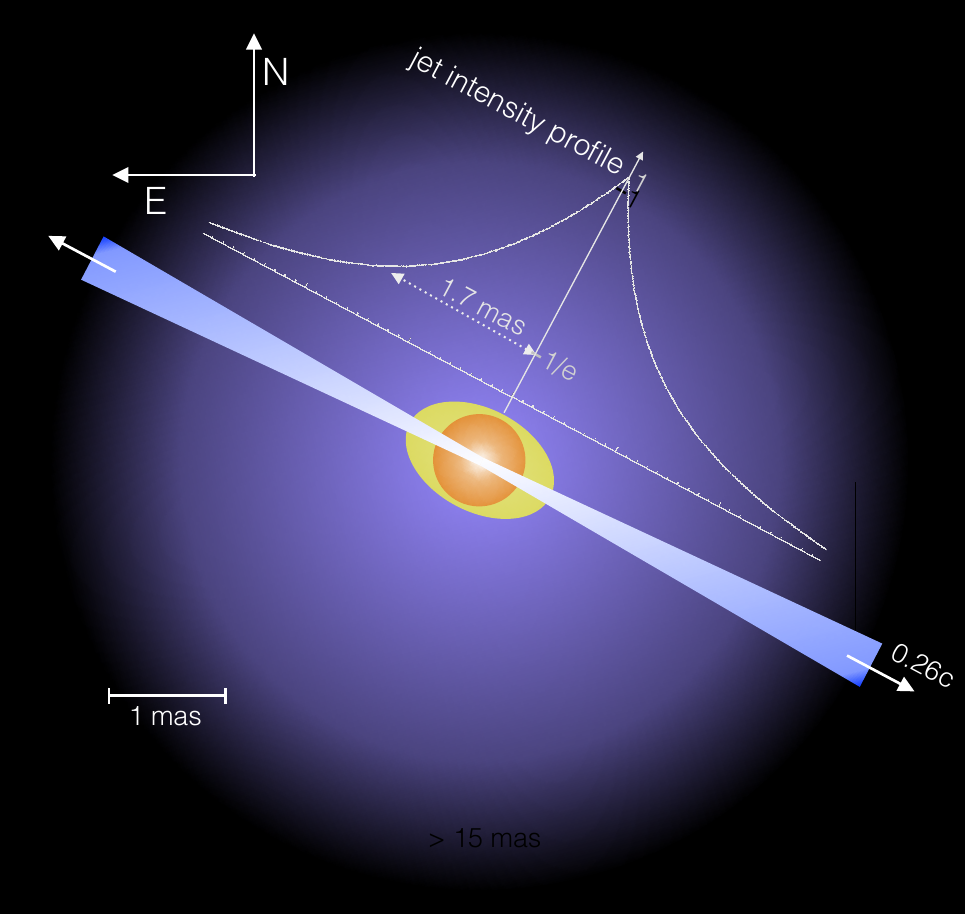}
\end{center}
 \caption{Sketch of the inner region of SS 433. Most (90\%) of the infrared continuum comes from a partially resolved central source of typical size $\sim$0.8 mas (central red area). The 10\% continuum flux left is produced by a completely resolved background on a larger scale ($>$15mas, background area). The \Brg line emitting region size is found to be $\sim$1 mas and is dominated by an East-West component, in the direction of the jet (yellow area). The measured position angle of the jet is $\sim$ 80$^{\circ}$. The jet profile is unresolved in the transverse direction ($<$1.2 mas). The intensity profile along the jet axis is best characterized by an emission peaking at the continuum position and decreasing exponentially on a scale of 1.7 mas.}
\label{sketch}
\end{figure}

Our GRAVITY observations of SS 433 allow us to resolve its accretion-ejection structure at sub-mas scale. A sketch summarizing our results is shown in Fig. \ref{sketch}. Most (90\%) of the infrared continuum comes from a partially resolved central source of typical size $\sim$0.8 mas potentially dominated by disk wind emission \citep{fuc06}. The 10\% continuum flux left could be produced by a completely resolved background on a larger scale ($>$15mas). This background shows no asymmetry (no closure phase is measurable in the continuum) that could suggest a link with the jets. On the other hand it is quite extended. Its exact origin is then unclear.
Jet lines are clearly present in the spectrum. The measured position angle of the jet is fully consistent with the  PA ($\sim$80$^{\circ}$) derived from the kinematic model at a precession phase $\psi=0.71$. The jet profile is unresolved in the transverse direction ($<$1.2 mas at 3$\sigma$), as expected since both optical and X-ray emission line widths indicate a jet opening angle $\theta_{j}\approx$ 1.5$^{\circ}$ \citep{bor87,mar02}. 
The intensity profile along the jet axis is best characterized in our data by an emission peaking at the continuum position and decreasing exponentially on a scale of 1.7 mas = 1.4$\times 10^{14}$\,cm. 

This profile for the jet intensity is certainly not unique. However, the quite strong decrease of the visibility amplitude across the jet lines for all the baselines as well as the similar phase behavior of the jet lines produced by the same jet component  strongly support a sharply decreasing two-sided intensity profile for the jet. 

During the $\sim$4h observation, the jet material traveling at 0.26c moves by $\sim$1.3 mas, i.e., of the order of the size of the jet structure. This implies that rapid flux decay must occur during propagation along the jet to explain the exponential decay of the jet intensity profile.
Interestingly, an exponential profile was also used by \cite{bor87} 
to fit the flux decay timescale of individual kinematic components in the H$\alpha$ profile.  They found an offset of the jet emission by 4$\times10^{14}$ cm (4.8 mas) from the core and with a larger spatial extent of 6.7$\times 10^{14}$\,cm (8.1 mas) (see also F04). A direct comparison is difficult however given the non-simultaneity of these observations and the known variability of the jet structure. Moreover, the NIR and optical lines may have different emissivity profiles (e.g. because of increased extinction in the optical compared to the NIR). 

Our observation unambiguously fixes the size of the jet infrared line emission region ($\sim 1.4\times 10^{14}$\,cm). Given the average jet line luminosities, this size is far too large for the (conical) jet to be entirely filled of emitting gas and a quite small clump filling factor ($\sim 10^{-5}$) is required (e.g. \citealt{dav80,beg80,pan97}). 
Clumping is likely to be due to thermal instabilities in a hot outflow, with the outflow initially driven by radiation pressure in the funnel of an accretion disk \citep{dav80}. X-ray line variability limits the hot outflow region to  $<4\times 10^{13}$\,cm \citep{mar13}. Our observation shows that the IR emission is displaced by less than 0.2 mas = 1.6$\times 10^{13}$\,cm from the central binary system, confirming that clumping originates early on in the hot outflow. It also suggests that line locking \citep{mil79}, if relevant to explain the amplitude and stability of the 0.26c jet velocity, operates on elements heavier than hydrogen to be efficient on this small scale \citep{sha86}. Models show that continuous heating of the clumps is required to explain optical emission on a scale of $10^{14}$\,cm \citep{bri88}. Future comparison of observed and theoretical intensity profiles may shed light on this heating process (\citealt{bri00}).

The stationary \Brg line shows a broad and double peaked profile and the interferometric observables suggest a geometry dominated by an East-West component, in the direction of the jet, like a (rather polar) disk wind. Then, both receding and approaching components must be present to explain the change of sign of the phases across the line. Taking the wavelength of the phase extrema as the corresponding blue and redshifted line produced by the wind, we infer a velocity of $\sim$600 km s$^{-1}$ and $\sim$2000 km s$^{-1}$ for the approaching and receding flow respectively. We also observe a similar dissymmetry  in the phases, which are significantly smaller in the blue part of the line.  Absorption effects could play a role here. Indeed at the precessional phase of the GRAVITY observation, the accretion disk has a near edge-on orientation and strong absorption from the wind could affect the line profile (like the P\,Cygni profile of \ion{He}{i}) especially in its blue part.  


The results presented here demonstrate the potential of spectro-interferometry to dissect the super-Eddington outflows and jets of SS 433. Additional insights will be gained in the future by monitoring this source at different precession and orbital phases with GRAVITY to obtain spectro-interferometric constrains on the stationary and jet line variability.


%

 %
 %


\section*{Acknowledgments}
This work is based on observations made with ESO Telescopes at the La Silla Paranal Observatory, programme ID 60.A-9102. It has been supported by a grant from LabEx OSUG@2020 (Investissements d'avenir -- ANR10LABX5) and in part by the National Science Foundation under Grant No. NSF PHY-1125915. It has made use of the Jean-Marie Mariotti Center \texttt{Aspro2} and \texttt{SearchCal} services\footnote{Available at http://www.jmmc.fr}. POP and GD acknowledge financial support from CNES and the French PNHE. JD was supported by a Sofja Kovalevskaja Award from the Humboldt Foundation of Germany. E.C. is supported by NASA through a Hubble Fellowship grant HST-HF2-51355 awarded by STScI, operated by AURA, Inc. We thank the ESO/VLTI team for their constant support. We also thank the technical, administrative and scientific staff of the participating institutes and the observatory for their extraordinary support during the development, installation and commissioning of GRAVITY.



\newpage
\begin{appendix}

\section{Jet lines spectral and interferometric signature modeling} 
\label{appA}
We decompose the contribution from the jet1 (approaching) and jet2 (receding) components separately  
as follows:
\begin{equation}
F_{\rm Norm}(\lambda) = F_{\rm jet1}(\lambda) + F_{\rm jet2}(\lambda) + 1
\end{equation}
\begin{equation}
F_{\rm jet1}(\lambda) =  \ion{He}{i}_{\rm jet1} +\ion{Br}{\gamma}_{\rm jet1}
\end{equation}
\begin{equation}
F_{\rm jet2}(\lambda) =  \ion{He}{i}_{\rm jet2} + \ion{Br}{\gamma}_{\rm jet2}+\ion{Br}{\delta}_{\rm jet2}
\end{equation}
where $ \ion{He}{i}_{\rm jet1/jet2}$, ${Br{\gamma}}_{\rm jet1/jet2}$ and  ${Br{\delta}}_{\rm jet2}$ are the flux ratios (fixed using the average spectrum) between the lines and the continuum. Assuming that the $Br{\gamma}$, \Brd and He I lines are emitted from the same regions, the interferometric differential observables are given then by Eq. \ref{eqmodel}. The characteristics of the different jet lines (redshift, FWHM, EW) are reported in Table \ref{tab1}.

\begin{table}[h!]
\begin{center}
\begin{tabular}{ccccc}
\hline
Name & Rest wavelength  & redshift & FWHM  & EW \\
 & ($\mu$m) &  & (km s$^{-1}$) & ($\AA$)\\
\hline
$\ion{Br}{\gamma}_{jet1}$ & 2.166 & 0.0132 & 1364 & 56\\
$\ion{Br}{\gamma}_{jet2}$ & 2.166 & 0.0624 & 1347 & 62\\
$\ion{Br}{\delta}_{jet2}$ & 1.944 & 0.0622 & 1510 & 59\\
$ \ion{He}{i}_{jet1}$ &2.057 & 0.0140 & 1429& 37\\
$ \ion{He}{i}_{jet2}$ &2.057 & 0.0631 & 1027& 44\\
\hline
\end{tabular}
\end{center}
\caption{Properties of each jet line. \label{tab1}}
\end{table}
\section{Normalized visibility amplitudes and phases of the six baselines}
\begin{figure*}[!b]
\begin{tabular}{c}
\includegraphics[width=0.4\columnwidth]{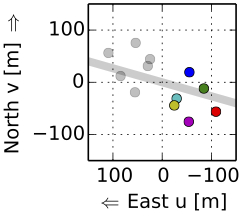} \\
\includegraphics[width=\textwidth,height=13cm]{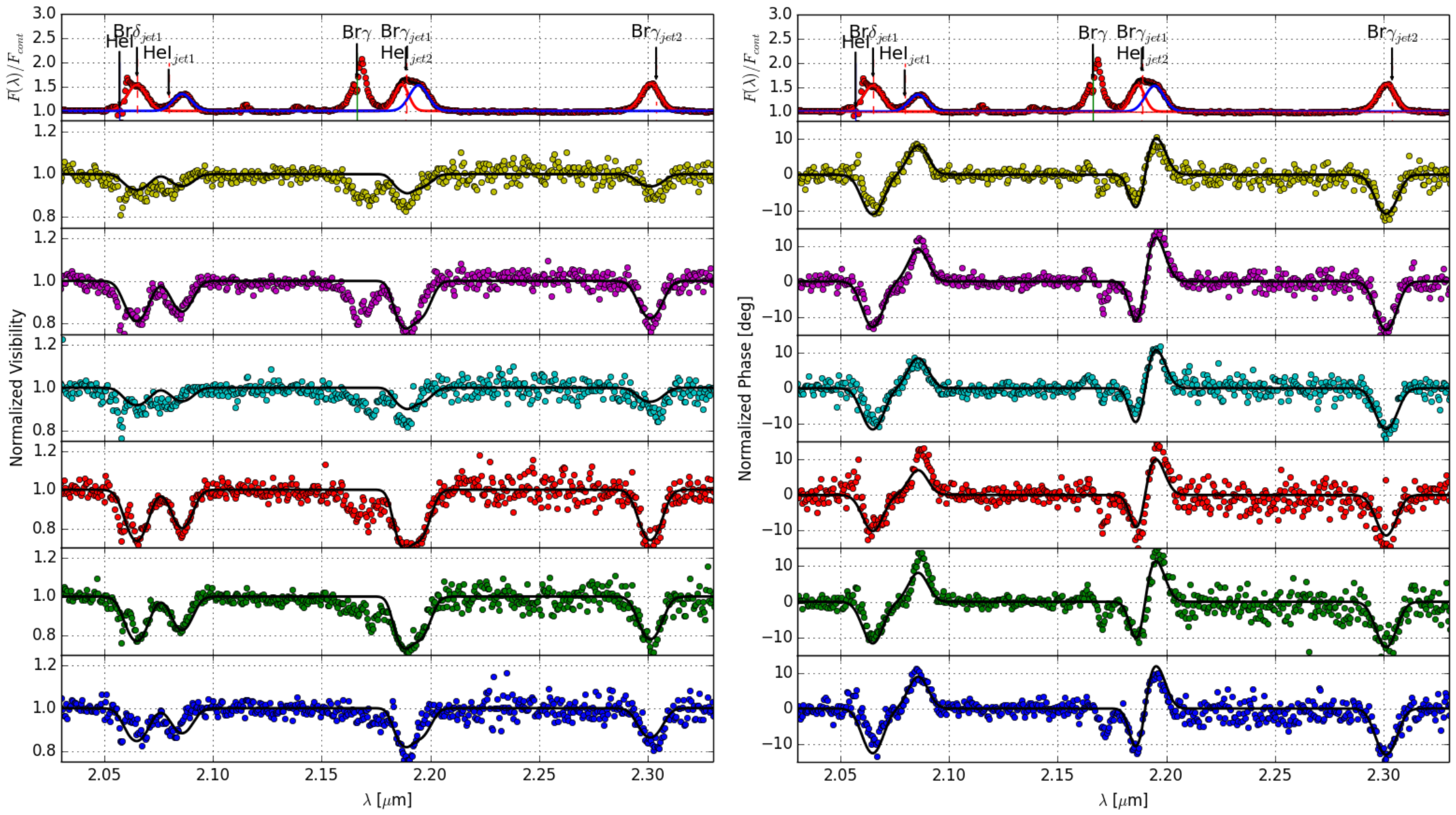} 
\end{tabular}
 \caption{Top) $uv$-plane at the time of the observation (average over the full exposure) with the different baselines indicated by colored points. The grey line represents the expected jet PA. Bottom) K-band GRAVITY spectrum (the solid blue/red line corresponds to the emission of the approaching/receding jet components) as well as the visibility amplitudes (left) and phases (right) for the 6 baselines. The solid line corresponds to the best fit model of the jet lines (see Sect. \ref{jetline}). The symbol colors correspond to the baseline colors indicated in the $uv$-plane.}
\label{figtotapp}
\end{figure*}

The $uv$-plane at the time of the observation, the K-band GRAVITY spectrum as well as the visibility amplitudes and phases for the 6 baselines are plotted in Fig. \ref{figtotapp}. 


\end{appendix}

\end{document}